\documentclass[12pt,a4paper]{article}

\usepackage[cp1251]{inputenc}
\usepackage[english]{babel}

\newtheorem{theorem}{Theorem}{\bf}{\it }
{\bf}{\it }

\title{Polynomial-time algorithm for determining the graph isomorphism}

\author{Anatoly D. Plotnikov}

\date{e-mail: a.plotnikov@list.ru}

\begin{document}
\maketitle

\begin{abstract}
We develop the methodology of positioning graph vertices relative to each other to solve the problem of determining isomorphism of two undirected graphs. Based on the position of the vertex in one of the graphs, it is determined the corresponding vertex in the other graph.

For the selected vertex of the undirected graph, we define the neighborhoods of the vertices. Next, we construct the auxiliary directed graph, spawned by the selected vertex. The vertices of the digraph are positioned by special characteristics --- vectors, which locate each vertex of the digraph relative the found neighborhoods. 

This enabled to develop the algorithm for determining graph isomorphism, the runing time of which is equal to $O(n^4)$.
\end{abstract}

\vspace{1pc}
{\bf MCS2000:} 05C85, 68Q17.


\section{Introduction}

Let $L_n$ is the set of all $n$-vertex undirected graphs without loops and multiple edges.

Let, further, there is a graph $G=(V,E)\in L_n$, where $V=\{v_1,v_2,\ldots,v_n\}$ is the set of graph vertices and $E=\{e_1,e_2,\ldots,e_m\}$ is the set of graph edges. Local degree $deg(v)$ of a vertex $v\in V_G$ is the number of edges which incident to the vertex $v$. Every graph $G\in L_n$ can be characterized by the vector $D_G=(deg(v_{i_1}),deg(v_{i_2}),\ldots,deg(v_{i_n}))$ of local vertex degrees, where $deg(v_i)\leq deg(v_j)$, if $i<j$.

The graphs $G=(V_G,E_G)$, $H=(V_H,E_H)\in L_n$ is called isomorphic if between their vertices there exists a one-to-one (bijective) correspondence $\varphi:$  $V_G\leftrightarrow V_H$ such that if $e_G=\{v,u\}\in E_G$ then the corresponding edge $e_H=\{\varphi(v),\varphi(u)\}\in E_H$, and conversely \cite{west}. The graph isomorphic problem consists in determining isomorphism graphs $G,H\in L_n$.

The problem of determining the isomorphism of two given undirected graphs is used to solve chemical problems, and to optimize programs. Effective (polynomial-time) algorithms for solving this problem were found for some narrow classes of graphs \cite{3,9}. However for the general case, effective methods for determining the isomorphism of graphs are not known.

The purpose of this article is to propose a polynomial-time algorithm for determining isomorphism of the connected undirected graphs.

\section{Algorithm bases}

We develop the methodology of positioning graph vertices relative to each other to solve the problem of determining isomorphism of two undirected graphs. Based on the position of the vertex in one of the graphs, it is determined the corresponding vertex in the other graph.

Consider the elements of our algorithm.

Let there be a graph $G\in L_n$. For the vertex $v\in V$ of the graph $G$ we define the concept of neighborhood of $k$th level $(0\leq k\leq n-1)$.

The neighborhood of 1st level of the vertex $v$ is the set of all graph vertices 
that are adjacent to the vertex $v$. In general, a neighborhood of $k$-th level is the set of all graph vertices that are adjacent to the vertices $(k-1)$th level. Such a neighborhood we denote $N_G^{(k)}(v)$. For convenience, we assume that the vertex $V$ forms a neighborhood of the zero level.

For the given vertex $v$ by means of the graph $G$ we construct an auxiliary directed graph $\vec G(v)$, spawned by the vertex $v\in V$, as follows.

The set of vertices belonging to one and the same neighborhood of the vertex $v$, form a line of graph vertices. Each line has the same number as the level of a neighborhood of the vertex $v$. Further, if in the initial graph $G$ the edge $\{v,u\}$ connects a vertex $v$ belonging to lines with a lower number than vertex $u$ then such an edge to replace by the arc $(v,u)$, outgoing from a vertex $v$ and incoming to the vertex $u$. If the edge of the graph connects vertices of one and the same line, this edge is replaced by two opposite arcs.

\begin{figure}[htbp]
\centering
\mbox{\input{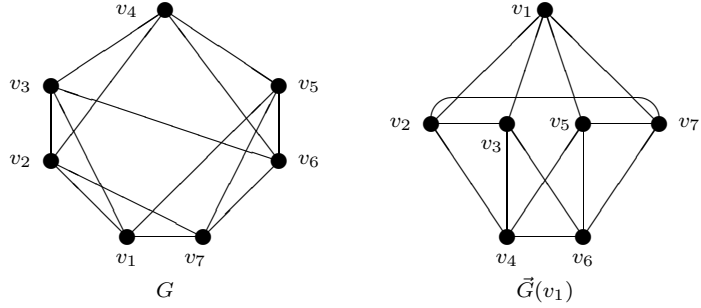}}
\caption{Graph $G$ and auxiliary digraph $\vec G(v_1)$}.
\label{grG}
\end{figure}

Fig. \ref{grG} presents the graph $G=(V,E)\in L_n$ and the auxiliary digraph $\vec G(v_1)$, spawned according to this graph by the vertex $v_1\in V$\footnote{The graph $G$ is borrowed from \cite{west}}. Here $N_G^{(0)}(v_1)=\{v_1\}$ is its own neighborhood (line) of the 0th level of the vertex $v_1$, the set of vertices $N_G^{(1)}(v_1)=\{v_2,v_3,v_6,v_7\}$ forms a neighborhood (line) of 1st level and, finally, the set $N_G^{(2)}(v_1)=\{v_3,v_6\}$ is a neighborhood of the 2nd level of the vertex $v_1$.

Each vertex $v$ of the auxiliary digraph $\vec G(v)$, we will characterize by two vectors $I_v$ and $O_v$.

Elements of vectors $I_v$ and $O_v$ are numbers. These numbers are the line numbers of vertices of the auxiliary digraph. Vector $I_v$ contains the line numbers of vertices, from which arcs of the digraph come into the vertex $v$, and the vector $O_v$ contains the line numbers of vertices that receive arcs from vertex $v$. If several arcs income into the vertex $v$ from one and the same line of digraph then the line number is repeated in the vector accordingly. Similarly, if several arcs from the vertex $v$ come  into one and the same line, this number is also repeated in the vector accordingly. The elements of the vectors $I_v$ and $O_v$ assume ordered in ascending order of value of numbers.

The vectors $I_v$ and $O_v$ will be called the characteristics of the vertex $v$, and $I_v$ is input and $O_v$ is output characteristics. Characteristics of the two vertices $v_1$, $v_2$ are equal, if their input and output characteristics are equal, respectively, i.e., if $I_{v_1}=I_{v_2}$ and $O_{v_1}=O_{v_2}$.
 
Find characteristics of the vertices of the auxiliary digraph $\vec G(v_1)$, shown in Fig. \ref{grG}.

\medskip

\begin{tabular}{|l|l|l|l|}
\hline
$I_{v_1}=\oslash$; & $I_{v_2}=(0,1,1)$; & $I_{v_3}=(0,1)$; & $I_{V_4}=(1,1,1,2)$;\\
$O_{v_1}=(1,1,1,1)$; & $O_{v_2}=(1,1,2)$; & $O_{v_3}=(1,2,2)$; & $O_{v_4}=(2)$;\\
\hline
$I_{v_5}=(0,1)$; & $I_{v_6}=(1,1,1,2)$; & $I_{v_7}=(0,1,1)$; & \\
$O_{v_5}=(1,2,2)$; & $O_{v_6}=(2)$; & $O_{v_7}=(1,1,2)$. & \\
\hline
\end{tabular}
\medskip


Auxiliary digraphs $\vec G(v)$ and $\vec H(u)$ is called positionally equivalent if the lines of graphs of the same level have the equal number of vertices, respectively, having equal characteristics.

\medskip

In general, the positionally equivalent digraphs have arcs connecting the vertices of the same level. This introduces an element of equivocation, i.e. in this case, we can not say that the positionally equivalent digraphs determine isomorphic graphs $G$ and $H$.
\medskip

A vertex $v\in V_G$ is called unique in digraph $\vec G(v)$ if doesn't exist other vertex with the characteristics equal to characteristics of the vertex $v$.

It is easy to see that the vertex $v_1$ is the unique vertex of the constructed digraph (see Fig. \ref{grG}).

\begin{theorem}
\label{tm1}
Let the graphs $G$ and $H$ are isomorphic. Let, further, the auxiliary digraphs 
$\vec G(v)$ and $\vec H(u)$ are the positionally equivalent, and each digraph has an unique vertex $v_i\in V_G$ and $u_j\in V_H$ such that $I_{v_i}=I_{u_j}$, $O_{v_i}=O_{u_j}$. Then between the vertices of digraphs there exists a bijective correspondence $\varphi$ such that $u_j=\varphi(v_i)$.
\end{theorem}

{\bf Proof}. Let the conditions of Theorem \ref{tm1} are satisfied. As the graphs $G$ and $H$ are isomorphic then between vertices of positionally equivalent digraphs $\vec G(v)$ and $\vec H(u)$, having equal characteristics, there is a bijective correspondence $\varphi$. Since the vertices $v_i$ and $u_j$ have the same unique positioning in the digraphs $\vec G(v)$ and $\vec H(u)$ then the relation $u_j=\varphi(v_i)$ is performed in any correspondence $\varphi$. In particular, that sort unique vertices are always vertices which spawn the positionally equivalent auxiliary digraphs.$\diamondsuit$.
\medskip

Note that in general case, the vertices of graphs $G$ and $H$ having the same local degree, spawn different auxiliary digraphs.

\begin{theorem}
\label{tm2}
Suppose there are two isomorphic graph $G=(V_G,E_G)$, $H=(V_H,E_H)\in L_n$. Suppose, further, for the subgraph $G_1\subseteq G$, induced by the vertex set $X\subseteq V_G$, the auxiliary directed graph $\vec G_1(x)$ has constructed. Then in the graph $H$ there exists a subgraph $H_1\subseteq H$, spawned by the set of vertices $Y\subseteq V_H$, such that its auxiliary digraph $\vec H_1(y)$ is positionally equivalent to the digraph $\vec G_1(x)$.
\end{theorem}

{\bf Proof}. The validity of the above statement follows from the definition of isomorphism of graphs, and the same sequence of construction of any of the auxiliary digraphs.$\diamondsuit$.
\medskip

\section{Algorithm}

Using the results of the previous section, we can propose several algorithms for determining the isomorphism of two given graphs $G,H\in L_n$. The simplest of them is as follows.

To determine the fact of isomorphism of graphs $G$ and $H$, we look for equally positioned vertex graphs.

Find the vertices $v\in V_G$ and $u\in V_H$ having positionally equivalent auxiliary digraphs in the graphs $G$ and $H$. Remove the found vertices from the graphs together with the incident edges and repeat the process until we have exhausted the list of vertices of graphs. If at some point in the graphs $G$ and $H$ cannot find a pair of vertices having positionally equivalent auxiliary digraphs, then stop the calculation, since the graphs are not isomorphic.

We describe the proposed algorithm in more detail.

Input of the algorithm: graphs $G=(V_G,E_G)$, $H=(V_H,E_H)\in L_n$, isomorphism of which must be determined, if it exists. We believe that these graphs have the same number of vertices and edges, as well as their vectors of local degrees $D_G$ and $D_H$ are equal.

Output of the algorithm: conclusion about the isomorphism of graphs $G$ and $H$.

\begin{list}{}{
\setlength{\topsep}{2mm}
\setlength{\itemsep}{0mm}
\setlength{\parsep}{1mm}
}
\item
{\bf Algorithm for determining isomorphism graphs.}

\item[{\it Step 1.}] Put $Q=G$, $S=H$, $N:=n$, $i:=1$, $j:=1$.

\item[{\it Step 2.}] Choose a vertex $v_i\in V_Q$ in the graph $Q$.

\item[{\it Step 3.}] Construct the auxiliary digraph $\vec Q(v_i)$ using the graph $Q$.

\item[{\it Step 4.}] Find the characteristics of vertices of the auxiliary graph $\vec Q(v_i)$.

\item[{\it Step 5.}] Choose the vertex $u_j\in V_S$ in the graph $S$.

\item[{\it Step 6.}] Construct the auxiliary digraph $\vec S(u_j)$ using the graph $S$.

\item[{\it Step 7.}] Find the characteristics of vertices of the auxiliary graph $\vec S(u_j)$.

\item[{\it Step 8.}] Compare the characteristics of the vertices of a digraphs $\vec Q(v_i)$ and $\vec S(u_j)$ in the neighborhoods of the vertices $v_i$ and $u_j$ of the same level.

\item[{\it Step 9.}] If the digraphs $\vec Q(v_i)$ and $\vec S(u_j)$ are positionally equivalent then put $V_Q:=V_Q\setminus \{v_i\}$, $V_S:=V_S\setminus \{u_j\}$, $N:=N-1$. Go to Step 11.

\item[{\it Step 10.}] If $j\leq N$ then put $j:=j+1$ and go to Step 5. Otherwise finish the calculations as the graphs $G$ and $H$ are not isomorphic.

\item[{\it Step 11.}] If $i\leq N$ then put $i:=i+1$, $j:=1$ and go to Step 2. Otherwise, stop the calculations. The graphs $G$ and $H$ are isomorphic.

\end{list}

\begin{theorem}
\label{tm5}
The algorithm for determining the graph isomorphism determines an isomorphism of the given graph if it exists.
\end{theorem}

{\bf Proof}. The Theorem \ref{tm1} is established that if graphs $G$ and $H$ are isomorphic then the pair of unique vertices $v$ and $u$ that spawn positionally equivalent digraphs $\vec G(v)$ and $\vec H(u)$ belong to some bijective mapping $\varphi$ of the vertices of this graph. Therefore, removal of vertices $v$ and $u$ of graphs $G$ and $H$ together with incident edges, leads to the construction of subgraphs of $G^{\prime}\subset G$ and $H^{\prime}\subset H$ which is also isomorphic. Therefore, the repetition of the above procedure will lead to the exhaustion of the list of vertices isomorphic graphs $G$ and $H$.

Now suppose that the graphs $G$ and $H$ are not isomorphic and the vertex $v\in V_G$, $u\in V_H$ spawn positionally equivalent digraphs $\vec G(v)$ and $\vec H(u)$. Then, obviously, in these digraphs there exist some subsets of vertices $X\subset V_G$ and $Y\subset V_H$, having corresponding equal characteristics, which spawn different (non-isomorphic) subgraphs. Here $v\in (V_G\setminus X)$, $u\in (V_H\setminus Y)$ and $Card(X)=Card(Y)$.

Removing sequentially from graph $Q$ and $S$ vertices, spawning the equivalent  auxiliary digraphs, we obtain non-isomorphic subgraphs. As a result, the algorithm terminates without possibility finding a pair of vertices with equivalent digraphs.$\diamondsuit$
\medskip

Unfortunately, the collection of pairs of vertices, determined by the proposed algorithm, as a whole, does not always determine the bijective mapping of the vertices in initial isomorphic graphs. This is the price for the simplicity of the algorithm.

\begin{theorem}
\label{tm6}
The running time of the algorithm for determining the graph isomorphism equal to $O(n^4)$.
\end{theorem}

{\bf Proof}. We determine the running time of the algorithm in steps 5--10.

Steps 5, 10 need to spend one time unit at each step.

Steps 6--8 require to spend $O(n^2)$ time units each.

Steps 9 require to spend $O(n)$ time units.

Therefore, $n$-multiple executing steps 5--10 requires to spend $O(n^3)$ time units.

Executing Steps 2--10 one times requires to spend $O(n^3)$ time units and executing $n$ times requires $O(n^4)$ time units.$\diamondsuit$

\section{Conclusion}

We have developed the efficient (polynomial-time) algorithms for determining isomorphism of undirected graphs. With that end in view, we used the methodology of positioning vertices concerning the selected vertex and its neighborhoods as the input and output characteristics of vertices. It allowed effectively to compare the structure of the given graphs.

Apparently, ideology of the positioning of vertices, developed in this study, can be used in solving the problem of finding a subgraph which is isomorphic to the given graph. Although in this case we should expect origin of essential search.

\newpage

{\bf \appendixname}
\vspace{1pc}

We illustrate the work of the proposed algorithm with an example.

\begin{figure}[htbp]
\centering
\mbox{\input{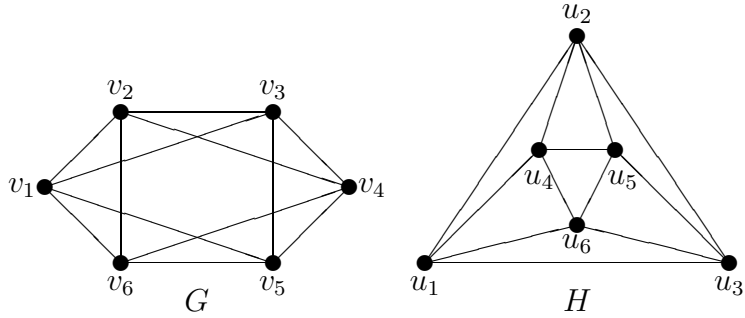}}
\caption{Initial graphs $G$ and $H$.}
\label{g1}
\end{figure}

Let two graph $G$, $H\in L_n$ are given, for which it is necessary to determine the isomorphism (see Fig. \ref{g1}).

The given graphs have an equal number of vertices, edges and vectors of degrees.

\begin{figure}[htbp]
\centering
\mbox{\input{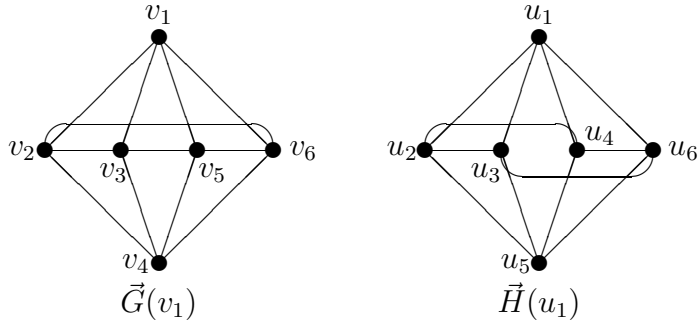}}
\caption{Auxiliary digraphs $\vec G(v_1)$ and $\vec H(u_1)$.}
\label{g2}
\end{figure}

Choose a vertex $v_1$ in the graph $G$ and construct the auxiliary digraph $\vec G(v_1)$ (see Fig. \ref{g2}).

Calculated characteristics of vertices of the constructed digraph. The results bring to the table.
\medskip

\begin{center}
\begin{tabular}{|l|l|l|}
\hline
$I_{v_1}=\oslash$; & $I_{v_2}=(0,1,1)$; & $I_{v_3}=(0,1,1)$;\\ 
$O_{v_1}=(1,1,1,1)$; & $O_{v_2}=(1,1,2)$; & $O_{v_3}=(1,1,2)$;\\
\hline
$I_{v_4}=(1,1,1,1)$; & $I_{v_5}=(0,1,1)$; & $I_{v_6}=(0,1,1)$;\\
$O_{v_4}=\oslash$; & $O_{v_5}=(1,1,2)$; & $O_{v_6}=(1,1,2)$;\\
\hline
\end{tabular}
\end{center}
\medskip

Choose a vertex $u_1$ in the graph $H$ and build an auxiliary digraph $\vec H(u_1)$ (see Fig. \ref{g2}).

The calculated characteristics of vertices of the newly constructed digraph. The results bring to the table.
\medskip

\begin{center}
\begin{tabular}{|l|l|l|}
\hline
$I_{u_1}=\oslash$; & $I_{u_2}=(0,1,1)$; & $I_{u_3}=(0,1,1)$;\\ 
$O_{u_1}=(1,1,1,1)$; & $O_{u_2}=(1,1,2)$; & $O_{u_3}=(1,1,2)$;\\
\hline
$I_{u_4}=(0,1,1)$; & $I_{u_5}=(1,1,1,1)$; & $I_{u_6}=(0,1,1)$;\\
$O_{u_4}=(1,1,2)$; & $O_{u_5}=\oslash$; & $O_{u_6}=(1,1,2)$;\\
\hline
\end{tabular}
\end{center}
\medskip

It is easy to see that the constructed auxiliary directed graph $\vec G(v_1)$ and $\vec H(u_1)$ positionally  equivalent. 

\begin{figure}[htbp]
\centering
\mbox{\input{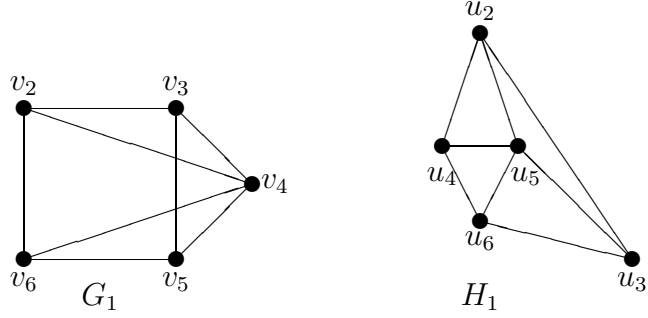}}
\caption{Subgraphs $G_1$ and $H_1$.}
\label{g3}
\end{figure}

Vertices $v_1$ and $u_1$ are removed from the initial graphs $G$ and $H$ respectively. We obtain graphs $G_1$ and $H_1$ (see Fig. \ref{g3}).

In the subgraph $G_1$, choose a vertex $v_2$ and build an auxiliary digraph $\vec G_1(v_2)$ (see Fig. \ref{g4}).

We calculate the characteristics of vertices of the auxiliary digraph $\vec G_1(v_2)$.
\medskip

\begin{center}
\begin{tabular}{|l|l|l|}
\hline
$I_{v_2}=\oslash$; & $I_{v_3}=(0,1)$; & $I_{v_4}=(0,1,1)$;\\ 
$O_{v_2}=(1,1,1)$; & $O_{v_3}=(1,2)$; & $O_{v_4}=(1,1,2)$;\\
\hline
$I_{v_5}=(1,1,1)$; & $I_{v_6}=(0,1)$; & \\
$O_{v_5}=\oslash$; & $O_{v_6}=(1,2)$; & \\
\hline
\end{tabular}
\end{center}
\medskip

In the subgraph of $H_1$, choose a vertex $u_2$ and build an auxiliary digraph $\vec H_1(u_2)$ (see Fig. \ref{g4}).

\begin{figure}[htbp]
\centering
\mbox{\input{g4.pic}}
\caption{Auxiliary digraphs $\vec G_1(v_2)$ and $\vec H_1(u_2)$.}
\label{g4}
\end{figure}

Calculate characteristics of vertices of the auxiliary digraph $\vec H_1(u_2)$.
\medskip

\begin{center}
\begin{tabular}{|l|l|l|}
\hline
$I_{u_2}=\oslash$; & $I_{u_3}=(0,1)$; & $I_{u_4}=(0,1)$;\\ 
$O_{u_2}=(1,1,1)$; & $O_{u_3}=(1,2)$; & $O_{u_4}=(1,2)$;\\
\hline
$I_{u_5}=(0,1,1)$; & $I_{u_6}=(1,1,1)$; & \\
$O_{u_5}=(1,1,2)$; & $O_{u_6}=\oslash$; & \\
\hline
\end{tabular}
\end{center}
\medskip

Again we find that the constructed auxiliary digraphs $\vec G(v_2)$ and $\vec H(u_2)$ are positionally equivalent.
 
\begin{figure}[htbp]
\centering
\mbox{\input{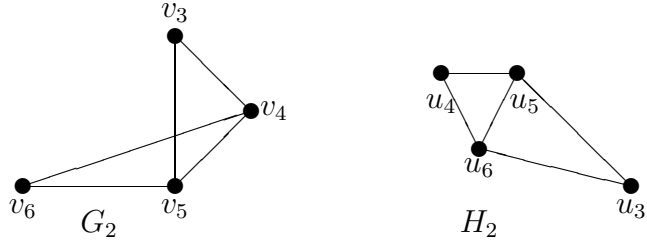}}
\caption{Subgraphs $G_2$ and $H_2$.}
\label{g5}
\end{figure}

Vertices $v_2$ and $u_2$ are removed from the subgraphs $G_1$ and $H_1$, respectively.
We obtain subgraphs $G_2$ and $H_2$ (see Fig. \ref{g5}).

\begin{figure}[htbp]
\centering
\mbox{\input{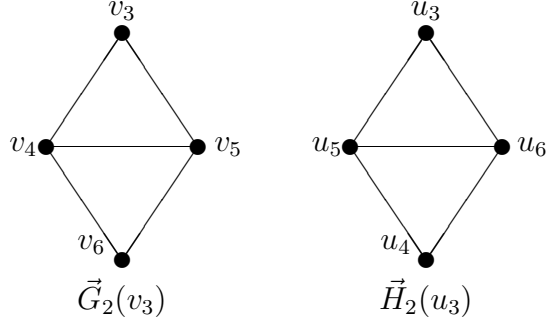}}
\caption{Auxiliary digraphs $\vec G_2(v_3)$ and $\vec H_2(u_3)$.}
\label{g6}
\end{figure}

In the graph $G_2$, choose a vertex $v_3$ and build an auxiliary digraph $\vec G_2(v_3)$ (see Fig. \ref{g6}).

Calculate characteristics of vertices of the auxiliary digraph $\vec G_2(v_3)$.
\medskip

\begin{center}
\begin{tabular}{|l|l|l|l|}
\hline
$I_{v_3}=\oslash$; & $I_{v_4}=(0,1)$; & $I_{v_5}=(0,1)$; & $I_{v_6}=(1,1)$;\\ 
$O_{v_3}=(1,1)$; & $O_{v_4}=(1,2)$; & $O_{v_5}=(1,2)$; & $O_{v_6}=\oslash$;\\
\hline
\end{tabular}
\end{center}
\medskip

Construct an auxiliary directed graph $\vec H_2(u_3)$ and find the characteristics of its vertices.
\medskip

\begin{center}
\begin{tabular}{|l|l|l|l|}
\hline
$I_{u_3}=\oslash$; & $I_{u_4}=(1,1)$; & $I_{u_5}=(0,1)$; & $I_{u_6}=(0,1)$;\\ 
$O_{u_3}=(1,1)$; & $O_{u_4}=\oslash$; & $O_{u_5}=(1,2)$; & $O_{u_6}=(1,2)$;\\
\hline
\end{tabular}
\end{center}
\medskip

Auxiliary digraphs $\vec G_2(v_3)$ and $\vec H_2(u_3)$ are positionally equivalent. 

\begin{figure}[htbp]
\centering
\mbox{\input{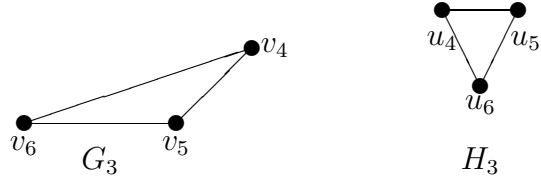}}
\caption{Subgraphs $G_3$ and $H_3$.}
\label{g7}
\end{figure}

Vertices $v_3$ and $u_3$ are removed from the graph $G_2$ and $H_2$ respectively.
We obtain graphs $G_3$ and $H_3$ (see Fig. \ref{g7}).

\begin{figure}[htbp]
\centering
\mbox{\input{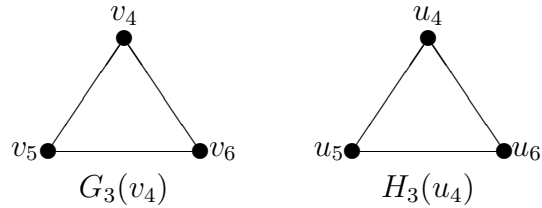}}
\caption{Auxiliary digraphs $\vec G_3(v_4)$ and $\vec H_3(u_4)$.}
\label{g8}
\end{figure}

In the graph $G_3$, choose a vertex $v_4$ and build an auxiliary digraph $\vec G_3(v_4)$ (see Fig. \ref{g8}).

We calculate the characteristics of the auxiliary graph $\vec G_3(v_4)$.
\medskip

\begin{center}
\begin{tabular}{|l|l|l|}
\hline
$I_{v_4}=\oslash$; & $I_{v_5}=(0,1)$; & $I_{v_6}=(0,1)$; \\ 
$O_{v_4}=(1,1)$; & $O_{v_5}=(1)$; & $O_{v_6}=(1)$; \\
\hline
\end{tabular}
\end{center}
\medskip

Construct an auxiliary digraph $\vec does h_3(u_4)$ and find the characteristics of its vertices.
\medskip

\begin{center}
\begin{tabular}{|l|l|l|}
\hline
$I_{u_4}=\oslash$; & $I_{u_5}=(0,1)$; & $I_{u_6}=(0,1)$; \\ 
$O_{u_4}=(1,1)$; & $O_{u_5}=(1)$; & $O_{u_6}=(1)$; \\
\hline
\end{tabular}
\end{center}
\medskip

Again we find that the constructed auxiliary digraphs $\vec G(v_4)$ and $\vec H(u_4)$ are positionally equivalent.

\begin{figure}[htbp]
\centering
\mbox{\input{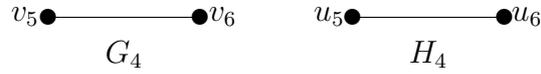}}
\caption{Subgraphs $G_4$ and $H_4$.}
\label{g9}
\end{figure}

Vertex $v_4$ and $u_4$ are removed from the graph $G_3$ and $does h_3$ respectively.
We obtain subgrapgs $G_4$ and $H_4$ (see Fig. \ref{g9}).

\begin{figure}[htbp]
\centering
\mbox{\input{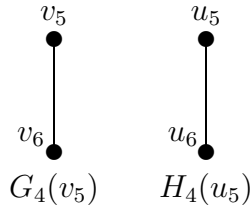}}
\caption{Auxiliary digraphs $\vec G_4(v_5)$ and $\vec H_4(u_5)$.}
\label{g10}
\end{figure}

In the graph of $G_4$, choose a vertex $v_5$ and build an auxiliary digraph $\vec G_4(v_5)$ (see Fig. \ref{g10}).

We calculate the characteristics of the auxiliary graph $\vec G_4(v_5)$.
\medskip

\begin{center}
\begin{tabular}{|l|l|}
\hline
$I_{v_5}=\oslash$; & $I_{v_6}=(0)$;  \\ 
$O_{v_5}=(1)$; & $O_{v_6}=\oslash$. \\
\hline
\end{tabular}
\end{center}
\medskip

Construct an auxiliary digraph $\vec H_4(u_5)$ (see Fig. \ref{g10}) and find the characteristics of its vertices.

\medskip

\begin{center}
\begin{tabular}{|l|l|}
\hline
$I_{u_5}=\oslash$; & $I_{u_6}=(0)$;  \\ 
$O_{u_5}=(1)$; & $O_{u_6}=\oslash$;  \\
\hline
\end{tabular}
\end{center}
\medskip

We find that auxiliary digraphs $\vec G_4(v_5)$ and $\vec H_4(u_5)$ are positionally equivalent.
 
Vertices $v_5$ and $u_5$ are removed from the subgraphs $G_4$ and $H_4$, respectively. We obtain one-vertex subgraphs $G_5$ and $H_5$. Auxiliary digraphs $\vec G_5(v_6)$ and $\vec H_5(u_6)$ contain one vertex and,of course, are the positionally equivalent.

Conclusion: the given graphs $G$ and $H$ are isomorphic.

\end{document}